\documentclass[final,5p,times,twocolumn]{elsarticle}
\usepackage[utf8]{inputenc}
\usepackage{booktabs}
\usepackage{tabulary}
\usepackage{longtable}
\usepackage{url}

\journal{arXiv}
\usepackage{xspace}
\newcommand{\micro}{micrometeorological \xspace}
\begin{document}

\begin{frontmatter}

\title{Constructing a Searchable Knowledge Repository for FAIR Climate Data}

\author[1]{Mark Roantree} \author[2]{Branislava Lalic}
\author[3]{Stevan Savic}
\author[3]{Dragan Milosevic}
\author[1]{Michael Scriney}

\address[1]{Insight Centre for Data Analytics, School of Computing, Dublin City University, Ireland\\}
\address[2]{Faculty of Agriculture, University of Novi Sad, Serbia.\\}
\address[3]{Climatology and Hydrology Research Centre, University of Novi Sad, Serbia.}

            




\begin{abstract}
The development of a knowledge repository for climate science data is a multidisciplinary effort between the domain experts (climate scientists), data engineers whos skills include design and building a knowledge repository, and machine learning researchers who provide expertise on data preparation tasks such as gap filling and advise on different machine learning models that can exploit this data.
One of the main goals of the CA20108 cost action is to develop a knowledge portal that is fully compliant with the FAIR principles for scientific data management. In the first year, a bespoke knowledge portal was developed to capture metadata for FAIR datasets. Its purpose was to provide detailed metadata descriptions for shareable \micro data using the WMO standard. While storing Network, Site and Sensor metadata locally, the system passes the actual data to Zenodo, receives back the DOI and thus, creates a permanent link between the Knowledge Portal and the storage platform Zenodo. While the user searches the Knowledge portal (metadata), results provide both detailed descriptions and links to data on the Zenodo platform. 

\end{abstract}




\end{frontmatter}


\section{Introduction and Background}
\label{sec:intro}
One of significant challenges of 21st century, climate change is accelerating \cite{IPCC2019}, together with the frequency and intensity of weather-related natural hazards and extreme events \cite{WEF2019}. A solid strategic plan for addressing weather and changing climate relies on high spatial density and good quality of site-specific
observations and measurements. This is crucial in order to obtain the necessary high-resolution numerical weather prediction, climate, agricultural, environmental and urban simulation models. This creates the need for adequate measurement methods and strategic planning for establishing application-oriented datasets, where some form of knowledge management can significantly contribute to, and set new and innovative standards.

Reliable and robust spatial and temporal data describing the environmental conditions acquired from
\micro data should provide a platform for the assessment and
modelling of both trends and effects of climate change (CC) and events such as adverse weather conditions on the environment across different ecosystems. In Europe, significant efforts have been made to centralise data from ground-based (at a synoptic scale) and satellite measurements, weather and climate simulations and by creating repositories such as COPERNICUS \cite{Coper2014} and ECMWF \cite{ECMWF2011}, which are available for public use. While these well established data sources are broadly employed in research, education and economics, what is missing from these repositories is \micro data: data addressing meteorological conditions of \emph{micro-environments}, that are both open and available to scientific researchers, stakeholders and other interested user groups.

Micrometeorological data is generally created as part of scientific projects and observational networks developed for different purposes, but they often \emph{languish} in reports, institutional data stores, or worse, personal computers. To address this failure to fully exploit this rich source of knowledge, the FAIRNESS Cost Action \cite{fairness2021} was funded to  establish a \micro knowledge share platform. Among its principal goals are: a repository of available and quality proven European (and beyond) \micro in situ datasets; the management of this data so that it complies to FAIR principles \cite{Wilkinson2016}; the normalisation of data quality and gap filing functions; to provide rich metadata descriptions for \micro data; and to make available, both rural and urban FAIR datasets.

\subsection{Cost Action Motivation}
Weather station networks are installed to monitor synoptic scale processes, as part of the Global Climate Observation System (GCOS). However, these networks are not sufficiently fine-grained to generate data about the state of the atmosphere at the micro-scale level. Current high-quality \micro databases focus on tower measurements and surface-atmosphere flux exchange \cite{Lappalainen2018, Pastorello2017} to provide information about the lower atmosphere. They are organised within permanently operating European (European Fluxes database cluster) or US (e.g. FLUXNET \cite{FLUXNET2020}) flux measurement networks and initiatives. However, spatial coverage is and will remain limited due to significant investment and maintenance costs. In terms of researchers, the target base for the FAIRNESS Cost Action is primarily, networks of Automated Weather Stations (AWSs) installed in rural, sub-urban and urban areas as part of various agrometeorology projects, such as pest and disease warning systems, forest, urban and environmental meteorology \cite{Lalic2020, Muller2013}. One of major weaknesses with existing projects is the lack of findability, accessibility, interoperability and reusability of the data: they fail to meet the FAIR Guide Principles in managing data and metadata \cite{Wilkinson2016}. 

\begin{figure}[h]
    \centering
    \includegraphics[width=8.5cm]{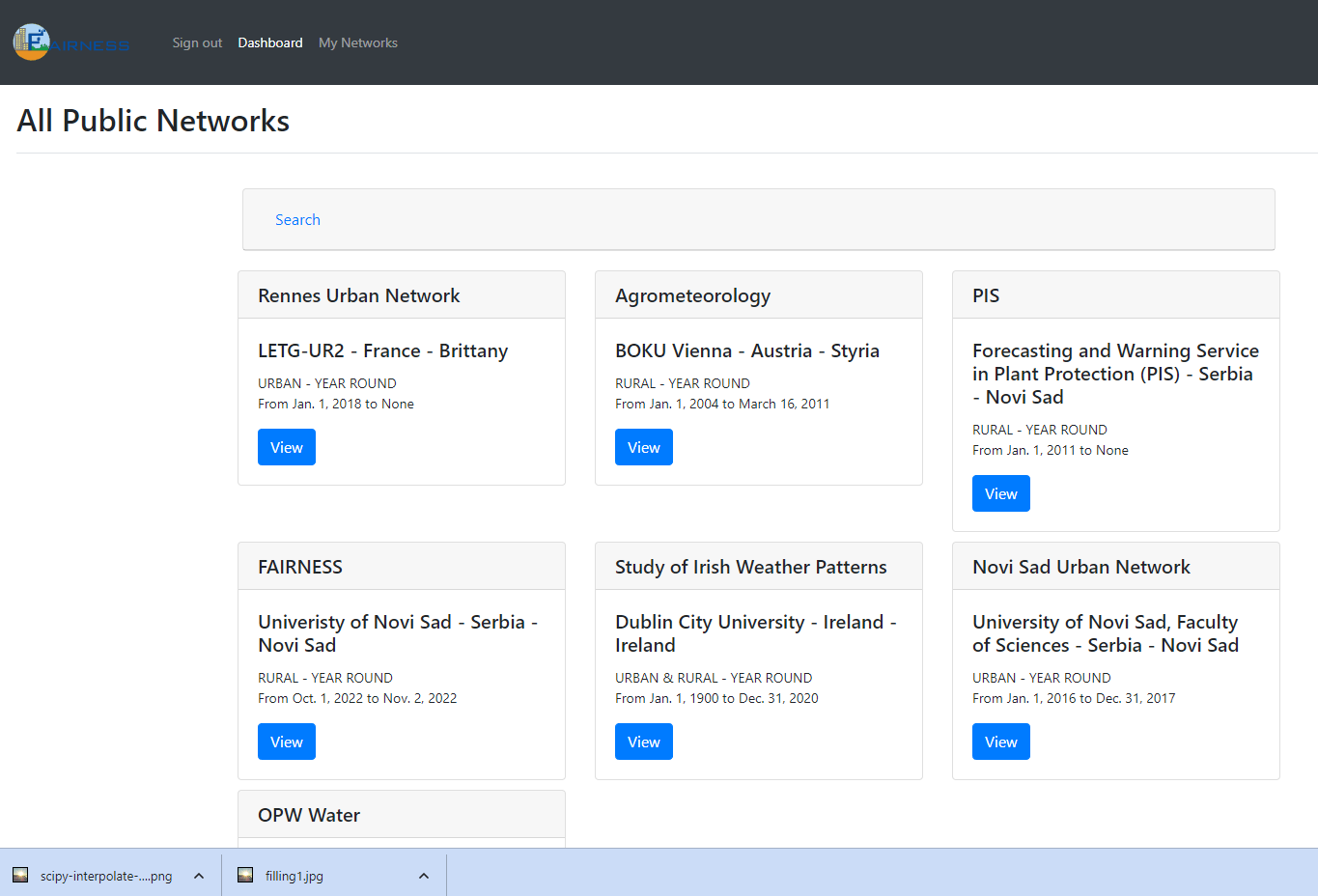}
    \caption{KSP Landing Page}
    \label{fig:kspfront}
\end{figure}

This paper describes the efforts of researchers within the CA20108 Cost Action \cite{fairness2021} to design and build a knowledge portal \cite{Loebbecke2012, McCarren2017} in which to store \micro metadata in order to make large volumes of available data FAIR compliant and where possible, openly available. Figure \ref{fig:kspfront} shows the landing for the current version of the FAIR Knowledge Portal where published network metadata is displayed. Potential benefits to a successful deployment of the knowledge portal are: better understanding of the current state of small-scale climate change and adverse weather effects in rural and urban areas; enhanced planning of field crop operations (irrigation), plant protection measures (spraying), improved and more effective forestation; enhanced planning of urban energy consumption; improved adaptation options for fighting carriers of dangerous viruses; organisation of more effective surveillance of invasive species such as Aedes albopictus.

\textbf{Paper Structure.}
In \S\ref{sec:fair}, we provide a brief overview of FAIR principals and reference some initial efforts on the application of FAIR metrics to an existing project;  
In \S\ref{sec:search}, we present a discussion on searching using the Knowledge Portal;
In \S\ref{sec:analytics}, a brief overview of system analytics are presented;
and finally in \S\ref{sec:conc}, the paper finishes with conclusions.

In the remainder of the paper, the term FAIRNESS is used to refer to the FAIRNESS CA20108 Cost Action and FKP refers to the FAIRNESS Knowledge Portal.

\section{Application of FAIR Metrics}
\label{sec:fair}
The main goal and objectives of FAIRNESS are fully focused on the open science concept as one of the main pillars of future scientific strategy and are in line with several mission areas. Thus, it is useful to understand and test the FAIR metrics.

\begin{figure}[h]
    \centering
    \includegraphics[width=8.5cm]{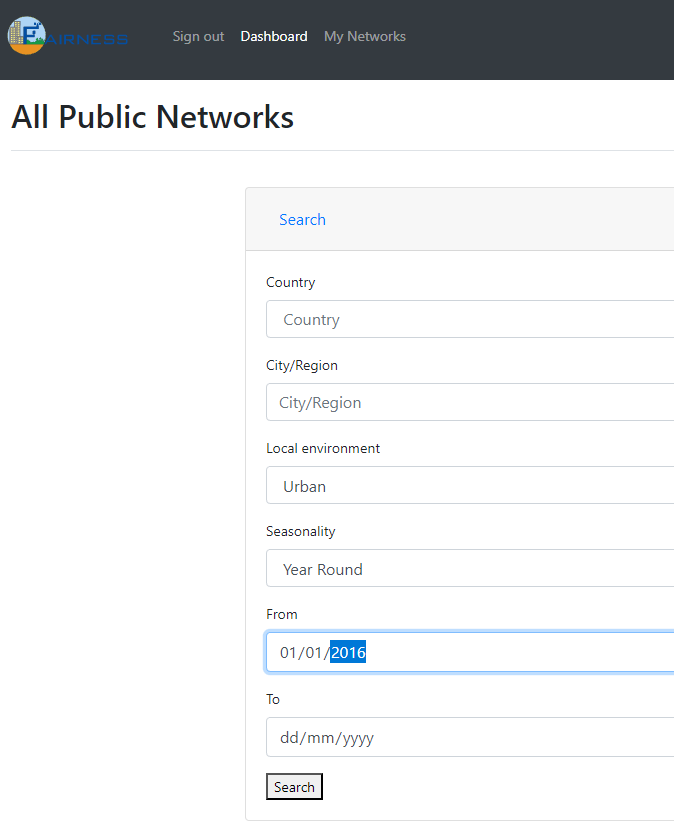}
    \caption{Findable Metadata: Initiate Search}
    \label{fig:findable}
\end{figure}

\begin{figure}[h]
    \centering
    \includegraphics[width=8.5cm]{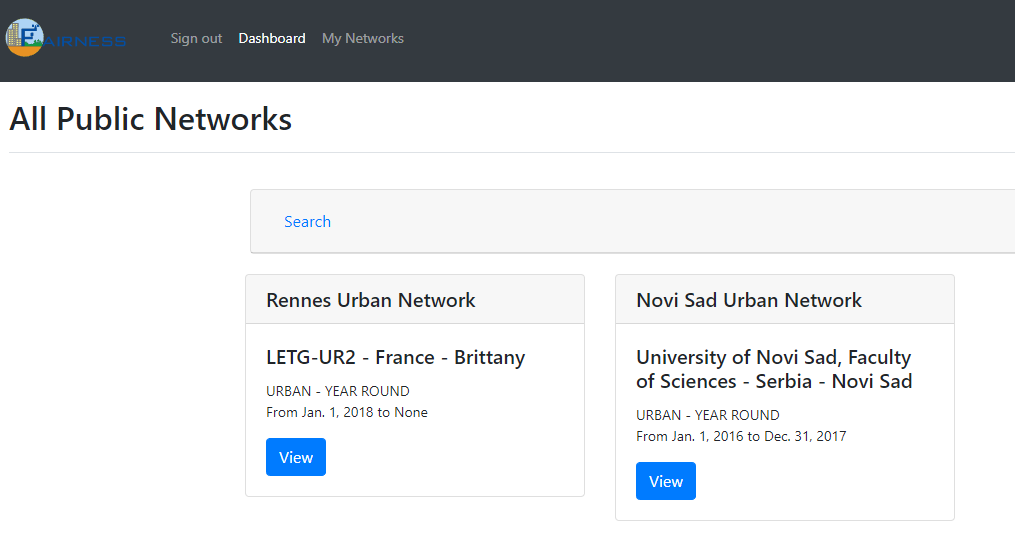}
    \caption{Search Results}
    \label{fig:search:res}
\end{figure}

We begin with a brief recap of the FAIR principles.	
\begin{itemize}
	\item \textbf{Findable.} Machine-readable metadata is required for automatic discovery of datasets and services. Using the KPG, a metadata description is supplied by the data owners for all micro-meteorological data shared on the system which subsequently drives the search engine, using keywords or network, site and sensor search terms. In figure \ref{fig:findable}, the method for retrieving metadata in the current version of the FKP is shown as a series of search fields: Country, Region, Local Environment, Seasonality and Dates. This is currently being expanded to include keyword searches across a broader range of metadata descriptions.

	\item \textbf{Accessible.} When datasets matching the search criteria have been identified (see figure \ref{fig:search:res}), network details are provided on the results page. Assuming data is freely accessible, Zenodo DOIs and links are provided for direct data access.
	
	\item \textbf{Interoperable.} Data interoperability means the ability to share and integrate data from different users and sources \cite{Scriney2019}. This can only happen if a standard (meta) data model is employed to describe data, an important concept which generally requires data engineering skills to deliver. In the FKP, the WMO guide provides the design and structure for metadata. 

	\item \textbf{Reusable.} To truly deliver reusability, metadata should be expressed in as detailed a manner as possible. In this way, data can be replicated and integrated according to different scientific requirements. While the FKP facilitates very detailed metadata descriptions, not all metadata is compulsory as it was accepted that in some cases, the overhead in providing this information can be very costly. 
\end{itemize}

It should be noted at this point that the original FAIR specification for data was relatively abstract, potentially leading to different interpretations. Some efforts have been made to further clarify the FAIR principles and even suggest ways to strengthen the original description \cite{Aquin2023}. This was achieved by examining existing open repositories such as EOSC and providing a deeper examination of how to assess some measure of FAIR for repositories. Interoperable data warehouses, including those constructed to use the XML standard such as \cite{Roantree2014}, provide an guide to developing an infrastructure for FAIR repositories that utilise a \emph{common data model}. This should form the core of any Knowledge Portal or these portals will struggle with the Interoperable principle for FAIR. During the development of the FKP, efforts were made to provide more detail on what constitutes FAIR and began by focusing on a standard data model (WMO GAMP \cite{GAMP2010}) to capture metadata descriptions. This design decision ensures an underlying support for FAIR in that it delivers the Interoperable principle for all FKP metadata and data.

In \cite{Lalic2022}, the authors present a case study which details the application of a FAIR test for \micro data. This extensive self assessment of PIS \micro comprised four phases: an initial self assessment to identify any lack of basic features using self assessment tools;
qualitative self assessment to test the quality of existing solutions and identify room for improvement; the identification of tools, methods and algorithms necessary to implement findings from the first 2 steps; and the execution and testing of one segment of a Forecasting and Reporting Service for Plant Protection in Serbia.

This paper reported the degree to which the forecasting and reporting data was FAIR compliant by testing against all 16 FAIR metrics (each of the 4 principles has 4 tests) with basic categories of Yes, No, Partial for each test. The \textbf{Findable} metrics are: a persistent identifier must be assigned to data (F1); there are rich metadata describing the data (F2); the metadata are online in a searchable resource (F3); and the metadata record specifies the persistent identifier. For these, the test dataset scored 2 \emph{Yes} and 2 \emph{Partial} responses. The \textbf{Accessible} test metrics are: following the persistent ID must lead to data or associated metadata (A1); the protocol by which data can be retrieved follows recognised standards (A2); the access procedure must include authentication and authorisation steps (A3); and metadata is accessible where possible even when data is not open or FAIR (A4). For Accessible metrics, the test dataset scored 2 \emph{Yes} and 2 \emph{Partial} responses. The \textbf{Interoperable} metrics are: Data is provided in a commonly understood and preferably open format (I1); metadata follows relevant standards (I2); controlled vocabularies, keywords, thesauri or ontologies are used where possible (I3); and qualified references and links are provided to all related data (I4).
For Interoperable metrics, the test dataset scored just one \emph{Yes} for I2 and 3 \emph{Partial} responses. The \textbf{Reusable} metrics are: data are accurate, well described with many relevant attributes (R1); data have a clear and accessible data usage license (R2); Clarity as to why and by whom data have been created and processed (R3); and data and metadata meet relevant domain standards (R4). This final test saw another 2 \emph{Yes} and 2 \emph{Partial} responses.

\begin{figure}[h]
    \centering
    \includegraphics[width=9cm]{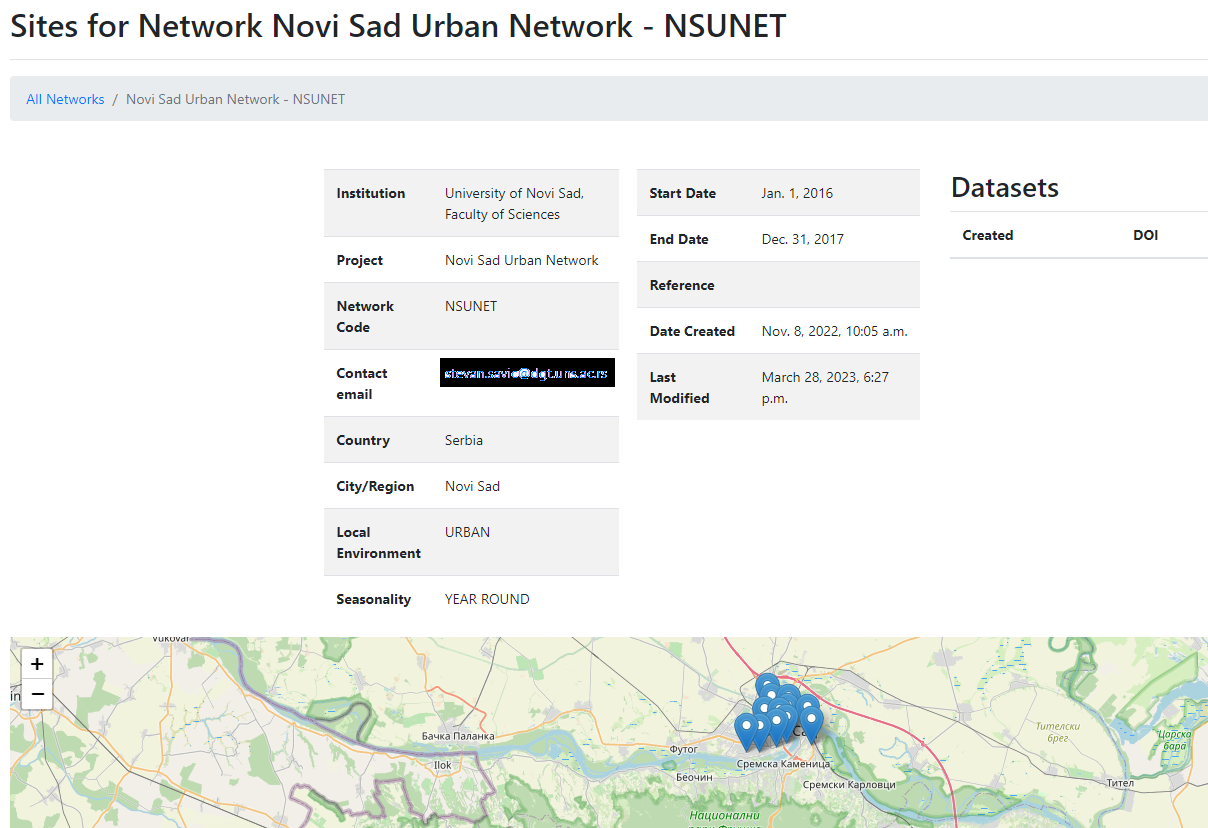}
    \caption{Examining Search Results}
    \label{fig:result:stevan}
\end{figure}

Conducted before the development of the FKP, it provided useful guidelines as to how to incorporate a FAIR test into the FKP and encouraged a focus on the \emph{Interoperability} metric as this is the most difficult metric on which to achieve full conformance.

\begin{figure*}[h]
    \centering
    \includegraphics[width=16cm]{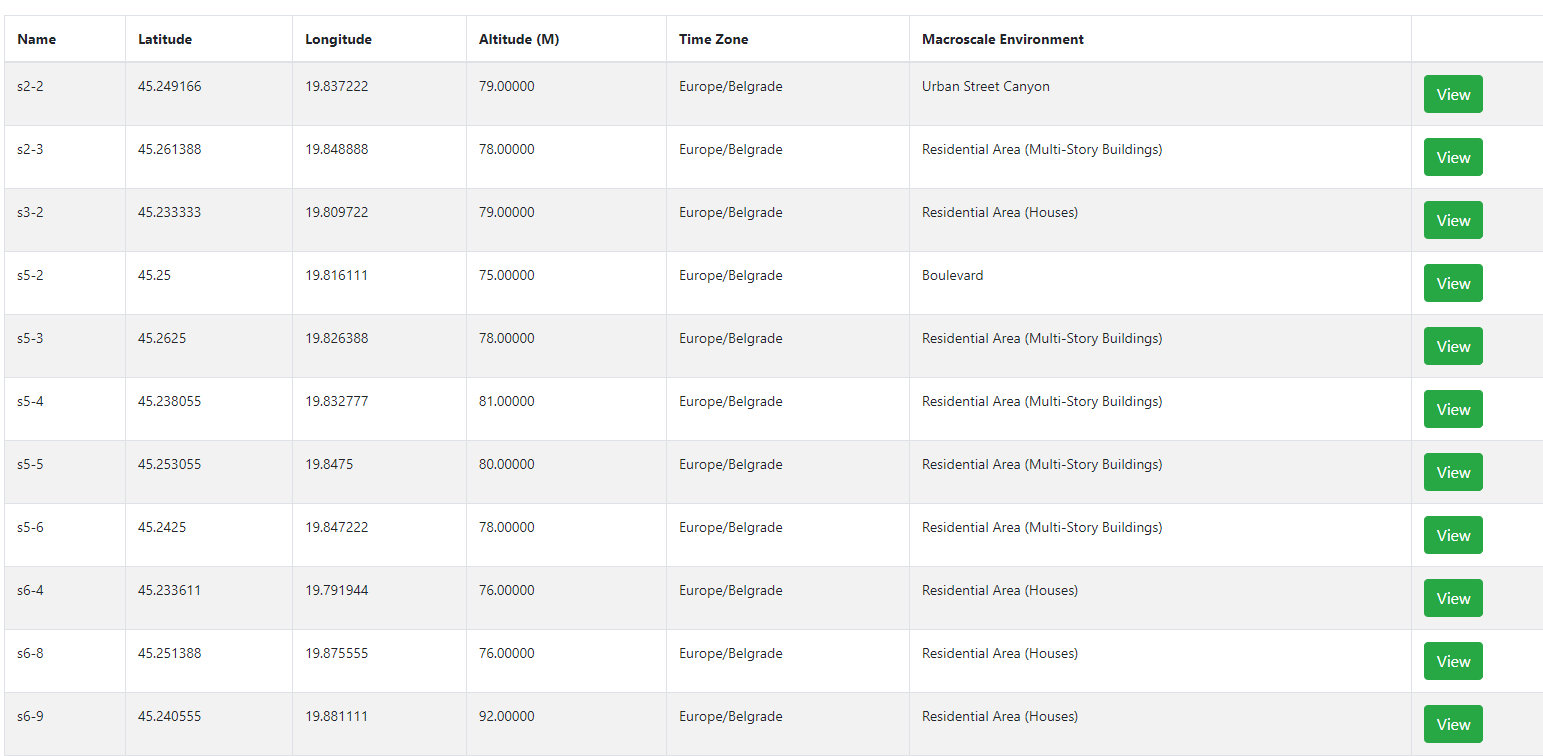}
    \caption{Examining Site Details}
    \label{fig:result:sites}
\end{figure*}

\section{Micro Climate Knowledge Portal}
\label{sec:search}
In this section, we use a medium sized \micro network to demonstrate features in the current version of the FKP. The research project which generated this dataset \cite{Secerov2019} resulted in the construction of the Novi Sad Urban Network (NSUNET), a collection of 28 remote stations and 2 servers built solely on opensource technologies. It's goals were the monitoring of climate incidents; the acquisition of long-term meteorological data from the urban area of Novi Sad; in addition to the early warning notification to the city emergency services of the current urban weather conditions. One of the major benefits of this type of deployment was its ability to operate at a low Internet service fee, and ensure high reliability and performance on low-budget hardware \cite{Secerov2019}. Here, we can use this network to show some of the search features where it was one of the 2 networks located during the search initiated in figure \ref{fig:findable} and displayed in the results in figure \ref{fig:search:res} as Novi Sad Urban Network.

The dataset stored on the FKP represents a subset of the data from the original experiment, and is openly available \cite{Savic2023}. Metadata describes air temperature data captured hourly from 12 of the urban sites in Novi Sad over a period of 2 years covering 2016 and 2017. There are 2 datasets in the collection: one dataset provides details about the 12 sites at which the temperature sensors are placed, while the second file contains air temperature data at the 12 locations. In all, the second dataset contains 17,544 instances of air temperature data. The temperature data has been cleaned and gap-filled so there are 24 measures at each site for each day. Once the target network is selected from the results page, figure \ref{fig:result:stevan} displays the network metadata together with a interactive map with pin locations for each of the 12 sites. The FKP portal supports a drilldown to view the sites as show in figure \ref{fig:result:sites}.

\begin{figure}[h]
    \centering
    \includegraphics[width=8.5cm]{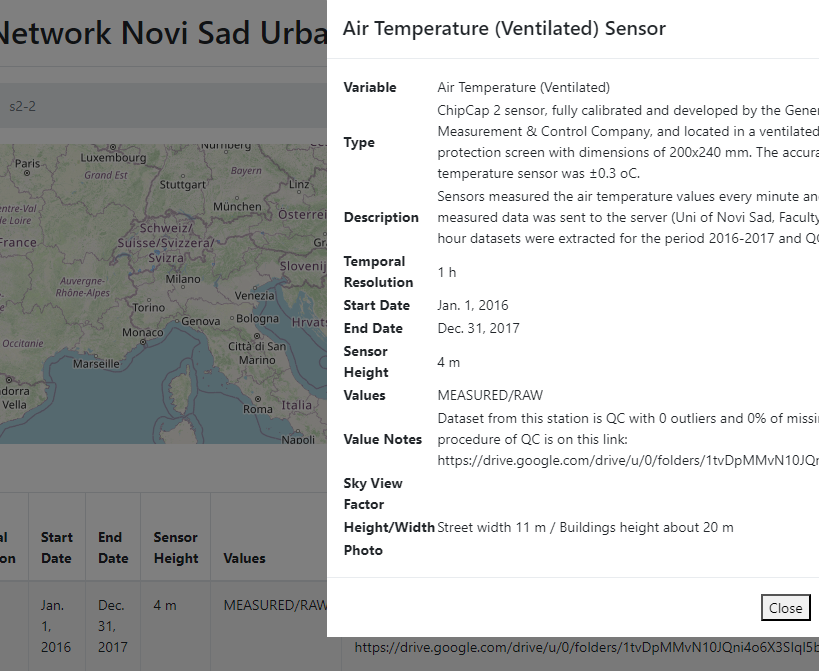}
    \caption{Specific Site Details}
    \label{fig:result:detail}
\end{figure}

Users can then drill down further to examine the metadata for any given site.  Figure \ref{fig:result:detail} displays the metadata captured for one of the sites in the Novi Sad network.

\section{KSP Analytics}
\label{sec:analytics}
Dimensional analytics are in place to monitor the volume and size of networks in the system, and provide some level of coverage in terms of network types, dates and geographic locations. By dimensional analysis, it means that FKP metadata can be analysed by: network, geographical location, dates, and urban/rural values. Drilldowns then support 2- or 3-dimensional analyses where dimensional metrics can be combined. For example, it is possible to count or locate the number of urban networks for a specific date range,for a specific country. 

Current metrics include: network count; average size of network (number of sites); dates and size of datasets per network/site; numbers and types of sensors in each site.

\section{Conclusions}
\label{sec:conc}
In this paper, we presented the FAIRNESS Knowledge Portal which was developed as part of the CA20108 Cost Action \cite{fairness2021}, using a step-by-step guide to demonstrate searching and browsing network \micro metadata. The current version of the FKP (v2.0) is open only to members of the Cost Action as we are still effectively in testing mode with the ongoing development of new features. This status is due to change at the end of the Cost Action when access will be open to the wider climate science community. 
Current plans include new Tools and Services to assess the quality of data, including the level of gaps and in some cases, machine learning tools will be provided to provide gap filling for datasets that meet certain criteria.

\pagebreak
\section*{Acknowledgements}
This work was conducted with the financial support of: the European Cooperation in Science and Technology (COST) Grant CA20108, Science Foundation Ireland under grant number SFI/12/RC/2289\_P2 and Ministry of Education, Science and Technological Development of the Republic of Serbia, Grant No. 451-03-47/2023-01/200117.

\section*{How to cite:}
Roantree, M., Lalić, B., Savić, S., Milošević, D., and Scriney, M. Constructing a Searchable Knowledge Repository for FAIR Climate Data, \\
EGU General Assembly2023, Vienna, Austria, 24–28 Apr 2023, EGU23-7786, \\https://doi.org/10.5194/egusphere-egu23-7786, 2023.

\end{document}